\documentclass[12pt]{article}

\renewcommand{\baselinestretch}{1.2}
\usepackage[margin=1.2in]{geometry}
\usepackage[T1]{fontenc}

\usepackage{fancyhdr}
\usepackage{hyperref}
\usepackage{xcolor}
\usepackage{amsmath,amssymb, amsfonts}
\usepackage[noadjust]{cite}
\usepackage{graphicx} 
\graphicspath{{./figures/}}
\usepackage{subfigure}
\usepackage{caption}
\usepackage{lipsum}
\usepackage[vlined,ruled]{algorithm2e}
\usepackage{tikz}
\usetikzlibrary{arrows.meta}
\usetikzlibrary{shapes.geometric,plotmarks,backgrounds,fit,calc,circuits.ee.IEC}
\usetikzlibrary{decorations.pathreplacing}

\newcommand{\transp}{\text{T}}

\DeclareMathOperator*{\rank}{rank}
\DeclareMathOperator{\im}{im}
\DeclareMathOperator{\supp}{supp}

\begin{document}
%
\title{Stabilizer Inactivation for Message-Passing Decoding of Quantum LDPC Codes}



\author{Julien Du Crest$^1$, \qquad Mehdi Mhalla$^2$, \qquad Valentin Savin$^3$\\[2mm]
    {\small $^1$\,Univ. Grenoble Alpes, Grenoble INP, LIG, F-38000 Grenoble, France}\\
    {\small $^2$\,Univ. Grenoble Alpes, CNRS, Grenoble INP, LIG, F-38000 Grenoble, France}\\
    {\small $^3$\,Univ. Grenoble Alpes, CEA-L\'eti, F-38054 Grenoble, France}\\
    {\small julien.du-crest@univ-grenoble-alpes.fr, mehdi.mhalla@univ-grenoble-alpes.fr, valentin.savin@cea.fr}
%
} 


%


\maketitle

\thispagestyle{fancy}

\begin{abstract}
We propose a  post-processing method for message-passing (MP) decoding of CSS quantum LDPC codes, called stabilizer-inactivation (SI). It relies on inactivating a set of qubits, supporting a check in the dual code, and then running the MP decoding again. This allows MP decoding to converge outside the inactivated set of qubits, while the error on these is determined by solving a small, constant size, linear system.
Compared to the state of the art post-processing method based on ordered statistics decoding (OSD), we show through numerical simulations that MP-SI outperforms MP-OSD  for different  quantum LDPC code constructions, different MP decoding algorithms, and different MP scheduling strategies, while having a significantly reduced complexity. We also provide
numerical evidence that SI post-processing may achieve a threshold on a family of generalized bicycle codes, of length varying from $126$ to $8190$ qubits.
%
\end{abstract}

\section{Introduction}
In a well-celebrated work~\cite{gallager1962low}, Gallager introduced the family of low-density parity-check (LDPC) codes, which became a staple of classical error correction over the past twenty years. They came equipped with iterative message-passing (MP) decoders, which is indisputably  the main reason of their success:  Not only MP decoders represented a low-complexity decoding approach (linear complexity per decoding iteration), but it was   later shown that LDPC codes can be optimized, so that their error correction thresholds under MP decoding  closely approach the theoretical Shannon limit~\cite{richardson2001capacity, richardson2001design}.

The quantum counterparts of LDPC codes, referred to as qLDPC codes, are Calderbank-Shor-Steane (CSS) quantum codes, defined by two orthogonal classical LDPC codes. qLDPC codes are likely to be the first practical codes used for fault-tolerant quantum computing~\cite{gottesman2014fault}, as they have low weight stabilizer group generators, which makes it possible to extract the error syndrome in a fault tolerant way. Moreover, the qLDPC family has been recently shown to yield good asymptotic codes, with linear minimum distance and constant rate~\cite{panteleev2021asymptotically, leverrier2022quantum}.  This augurs for practical constructions with increased error correction capacity, or reduced qubit overhead.


However, if decoding a qLDPC code boils down to decoding the two constituent classical LDPC codes (assuming separate decoding of $X$ and $Z$ errors), their orthogonality translates into a so-called degeneracy property, which causes a significant degradation of the MP decoding performance \cite{poulin2008iterative}. Thus, decoding solutions are usually devised on a case-by-case basis, being distinguished by their scope and extent. Heuristic techniques to partially overcome the degeneracy of qLDPC codes have been proposed in~\cite{poulin2008iterative}. Reweighted belief-propagation decoding has also been proposed in~\cite{babar2015fifteen} (and references therein), and more recently, neural belief-propagation decoding has been proposed in~\cite{liu2019neural}.  For hypergraph-product
LDPC and quantum expander codes, a linear-time decoding algorithm has been proposed in~\cite{leverrier2015quantum, fawzi2018efficient}, decoding up to a constant fraction of the minimum distance, by exploiting the expanding properties of the LDPC code. 
 Recently, Panteleev et Kalachev proposed a new decoding  approach, combining MP  with an ordered statistics decoding (OSD) post-processing step~\cite{panteleev2021degenerate}. 
 It stands currently as the most efficient decoding solution applying generally across a large spectrum of qLDPC codes~\cite{roffe2020decoding}. (Since the first version of this paper, linear time decoders for linear minimum distance qLDPC codes, decoding adversarial errors of linear weight, have been proposed in~\cite{gu2022efficient, leverrier2022efficient, dinur2022good}).

Here, we continue the idea of a general purpose decoder, in which MP decoding is combined with a post-processing step. We call our post-processing step {\em stabilizer inactivation} (SI). Assuming we are decoding for some type of errors, say $X$-errors, the idea of the SI post-processing step is to {\em inactivate} the qubits in the support of some $X$-check\footnote{A {\em check} is a generator of the stabilizer group, corresponding to a row of the  check matrix. Note that $Z$-checks are used to decode $X$-errors.}, and then run the MP decoding again. Inactivating these qubits  means that we take them out from the MP decoding process (which happens if we set to zero the corresponding log-likelihood ratio values, input to the MP decoder). If the inactivated $X$-check\footnote{By a slight abuse of language, we shall sometimes say that the $X$-check is inactivated (instead of the qubits in its support).} is carefully chosen, the MP decoding converges outside its support. All that remains is to determine the error on the support of the inactivated $X$-check, which amounts to solving a small linear system. We show that the MP-SI outperforms the MP-OSD decoder for different  qLDPC code constructions, different MP decoding algorithms, and different MP scheduling strategies, while having a significantly reduced complexity.

%

\section{Preliminaries}

\subsection{qLDPC Codes}
qLDPC codes are CSS quantum codes, defined by two classical LDPC codes with parity check matrices $H_X$ and $H_Z$, such that $H_X H_Z^\transp = 0$. The rows of $H_X$ and $H_Z$ are referred to as $X$-checks and $Z$-checks, respectively, and they generate the $X$-type and $Z$-type stabilizer groups, namely $\im H_X^\transp$ and $\im H_Z^\transp$. Here and henceforth, an element $g$ of the stabilizer group is identified to the binary vector indicating the qubits acted on non-trivially by $g$. 
 For a code of length $n$ qubits, $H_X$ and $H_Z$ have $n$ columns, while the number of encoded logical qubits is given by $k = n - \rank(H_X) - \rank(H_Z) = \dim \left( \ker H_X / \im H_Z^\transp \right) = \dim \left( \ker H_Z / \im H_X^\transp \right)$.  

The  condition $H_X H_Z^\transp = 0$ implies that any $Z$-check is a codeword of $\mathcal{C}_X = \ker H_X$, and any $X$-check is a codeword of $\mathcal{C}_Z = \ker H_Z$. Hence, $\mathcal{C}_X$ and $\mathcal{C}_Z$  have poor classical minimum distance, equal to $\mathcal{O}(1)$ as a function of $n$.  However, since the elements of the stabilizer group act trivially on any code state, the quantum  minimum distance is defined as $d = \min(d_X, d_Z)$, where $d_X = \min \left\{|v|, v \in \ker H_X \setminus \im H_Z^\transp\right\}$, and similarly, $d_Z = \min \left\{|v|, v \in \ker H_Z \setminus \im H_X^\transp\right\}$, and $|v|$ denotes the weight (number of $1$'s) of the binary vector $v$. As a function of $n$, the quantum minimum distance of qLDPC codes  varies from $d = \mathcal{O}(\sqrt{n})$ for the hypergraph-product (HP) construction  proposed in~\cite{tillich2013quantum} to $d = \mathcal{O}(n)$ for the lifted-product (LP) construction over non-abelian groups in~\cite{panteleev2021asymptotically}. Other constructions of qLDPC codes, with minimum distance in between the above bounds, have also been recently proposed in the literature~\cite{hastings2020fiber, panteleev2020quantum, breuckmann2020balanced}.

\subsection{Pauli Errors and Decoding}
\label{subsec:pauli_errors}
We assume that  encoded qubits are affected by  Pauli errors, where each qubit is independently acted on by a Pauli $I, X, Y, Z$ error, with probability $p_I, p_X, p_Y, p_Z$. The total error probability is denoted by $p = 1-p_I = p_X + p_Y + p_Z$. Since $Y=iXZ$, we only need to correct for $X$ and $Z$ errors.  
For simplicity, we shall assume that $X$ and $Z$ errors are decoded independently\footnote{Alternatively, one may decode first one type of error, say the $X$-error, then decode the $Z$-error conditional on the decoded $X$-error, in which case the $Z$-error channel model becomes a mixture of two BSCs.}, in which case the $X$-error channel model is a classical binary symmetric channel (BSC), with error probability $\varepsilon_X = p_X + p_Y$, and the $Z$-error channel model is a BSC, with error probability $\varepsilon_Z = p_Z + p_Y$.   

In the sequel, we shall only discuss the case of $X$-errors, since a similar discussion applies to $Z$-errors. We also identify an $X$-error to a binary vector $e_X$ of length $n$, where $1$ entries indicate the qubits on which an $X$ error occurred.  The error $e_X$ produces a syndrome $s_X := H_Z e_X$, which can be observed by measuring the $Z$-checks of the qLDPC code (here, we shall assume that the syndrome measurement is fault free). 

The task of the decoder is to determine an estimate $\hat{e}_X$ of the error $e_X$, given the observed syndrome $s_X$. Note that the same syndrome would have been produced by any other error $e'_X$, such that $e'_X + e_X\in \ker H_Z$. Decoding is successful if $\hat{e}_X + e_X\in \im H_X^\transp$, that is, the estimate and the true error differ by an element of the $X$-type stabilizer group. Indeed, such a mismatch is not problematic, since elements of $\im H_X^\transp$ acts trivially on any code state. Elements of $\im H_X^\transp$ are sometimes referred to as {\em degenerate errors}~\cite{poulin2008iterative}, since they need not (and actually cannot) be corrected. By  definition, a qLDPC code has many low-weight degenerate errors (any $X$-check is a degenerate error).

%

\subsection{MP Decoders}
We adopt the terminology related to the bipartite (Tanner) graph representation of classical LDPC codes~\cite{tanner1981recursive}. The bipartite graph has vertices referred to as bit-nodes and check-nodes, and its adjacency matrix is given by the parity check matrix of the code. An MP decoder is an iterative decoding algorithm  exchanging extrinsic messages along the edges of the bipartite graph, which can be described as follows~\cite{savin2014ldpc}:

\noindent (i) The input consists of the observed syndrome $s_X$, as well as an {\em a priori soft information} for each bit-node, usually given in the form of an log-likelihood ratio (LLR) value. Assuming the BSC model from the previous section, the a priori LLR is the same for all bit-nodes $i=1,\dots,n$, and given by 
\begin{equation}
\label{eq:a-priori-LLRs}
\gamma_i := \log\frac{\Pr(E_X(i) = 0)}{\Pr(E_X(i) = 1)}
 =  \log \frac{1-\varepsilon_X}{\varepsilon_X},
\end{equation}
where the unknown error $e_X$ is seen as a realization of a random variable, denoted here and in the sequel by $E_X$.

\noindent  (ii) Extrinsic messages are exchanged between bit-nodes and check-nodes, in several rounds, also referred to as decoding iterations. At each iteration, an {\em updated soft information} (posterior LLR\footnote{We  use (updated) ``soft information'' rather than ``posterior LLR'', since the later conveys a Bayesian inference meaning, which is only exact for the Sum-Product decoding on acyclic  graphs, as explained later in the text.}) value $\tilde{\gamma}_i$ is  computed for each each bit-node $i=1,\dots,n$, based on which an error estimate is made: 
\begin{equation}
\label{eq:hard-decision}
\hat{e}_X(i) = 0, \text{ if } \tilde{\gamma}_i \geq 0, \text{ and } \hat{e}_X(i) = 1, \text{ otherwise} 
\end{equation}

\noindent  (iii) Decoder stops if either $H_Z \hat{e}_X = s_X$, or a maximum (predetermined) number of decoding iterations is reached. It outputs the current error estimate $\hat{e}_X$ (hard decision), and possibly the corresponding soft information values (soft decision).


One may distinguish different MP decoding algorithms, the most well-known being the Sum-Product (SP, also known as Belief-Propagation\footnote{Many authors confuse BP with the general MP principle. We shall use SP terminology in this paper to avoid any possible confusion.}, BP~\cite{pearl1982bp}) and the Min-Sum (MS) algorithms.
On cycle free graphs, SP implements maximum a posteriori (MAP) decoding, while MS implements maximum-likelihood (ML) decoding. For graphs with cycles,  computation trees provide a way to describe SP and MS decoders as an either MAP or ML decision process on a cycle-free graph~\cite{wiberg1996codes}. 
%
%
Compared to SP, MS may present several practical advantages: it has a lower computational complexity (only requires additions and comparisons), and under the BSC model assumption, it does not need an a priori knowledge of the channel error probability $\varepsilon_X$. Hence, for the MS decoder,  the a priori input LLRs can be initialized to any constant value, $\gamma_i = \gamma, \forall i = 1,\dots,n$, without  modifying the decoder output. However, to avoid computation instability issues, $\gamma=1$ should be used\footnote{We warn the reader that computation instability may actually improve the MS decoding performance, depending on the $\gamma$ value, and this even for double-precision floating-point ($64$ bits) implementations. While for classical LDPC codes this impact is usually negligible, for qLDPC codes we have observed significant differences.  We believe this is due to phenomena related to the code degeneracy. See also the random perturbation technique  in~\cite{poulin2008iterative}.}.


MP decoders may use different scheduling strategies, determining the order in which bit-node and check-node messages are updated. 
 The conventional assumption is that at each iteration, all bit-nodes and subsequently all check-nodes pass new messages to their neighbors, which is known as parallel or flooding scheduling. A different approach, known as serial scheduling, is to process check-nodes  sequentially\footnote{We discuss here {\em horizontal} scheduling strategies, applying to check-nodes. Similar {\em vertical} scheduling strategies apply to bit-nodes.}, with an immediate propagation of their messages to the neighbor bit-nodes.  The main advantage of the serial schedule is that it propagates information faster and converges in about half the number of iterations compared to the flooding schedule~\cite{sharon2007efficient}. The serial scheduling can be also partly
parallelized, by processing sequentially layers (groups) of check-nodes, while processing in parallel check-nodes within the same layer. This scheduling technique is known as layered scheduling. In case that any bit-node is connected at most once to each check-node layer,  serial and layered scheduling exhibit {\em strictly the same} decoding performance and convergence speed. The choice of a scheduling strategy is usually related to a specific decoder architecture implemented in hardware. Partly parallel architecture implementing a layered scheduling are widely used in practical implementations~\cite{ boutillon2014hardware}, but ultra-high throughput hardware implementations are based on unrolled fully-parallel architectures~\cite{schlafer2013new}, implementing a flooding scheduling.

\subsection{OSD Post-Processing}

The MP-OSD decoder proposed in~\cite{panteleev2021degenerate} combines  MP decoding with an OSD post-processing step, in case the former fails to find a hard decision estimate  $\hat{e}_X$ satisfying the given syndrome $s_X$. The post-processing step utilizes the soft decision outputted by the MP decoder.  In its most basic form (OSD-$0$), bit-nodes are  sorted according to their {\em reliability} (that is, absolute value of  the corresponding soft decision), and a number of $\rank{H_Z}$ least reliable bit-nodes are determined, such that the corresponding columns of $H_Z$ are linearly independent. The hard decision estimates of these bit-nodes are discarded, and they are then re-estimated by solving the linear system determined by the corresponding columns (for the remaining, more reliable bit-nodes, the hard decision estimates outputted by the MP decoder are kept). Hence, OSD always outputs a valid solution $\hat{e}_X$ of the system $H_Z e_X = s_X$. In~\cite{panteleev2021degenerate}, MP-OSD has been shown to produce effective results,  when the MP decoder is a normalized version of the MS (NMS) using a serial scheduling.


%

\section{Stabilizer Inactivation Post-Processing}


In the  following, we will only discuss the correction of $X$ errors, but similar arguments apply to $Z$-errors.


\subsection{Stabilizer-Splitting Errors}

\begin{figure}[!t]
\centering
\includegraphics[width=.4\linewidth]{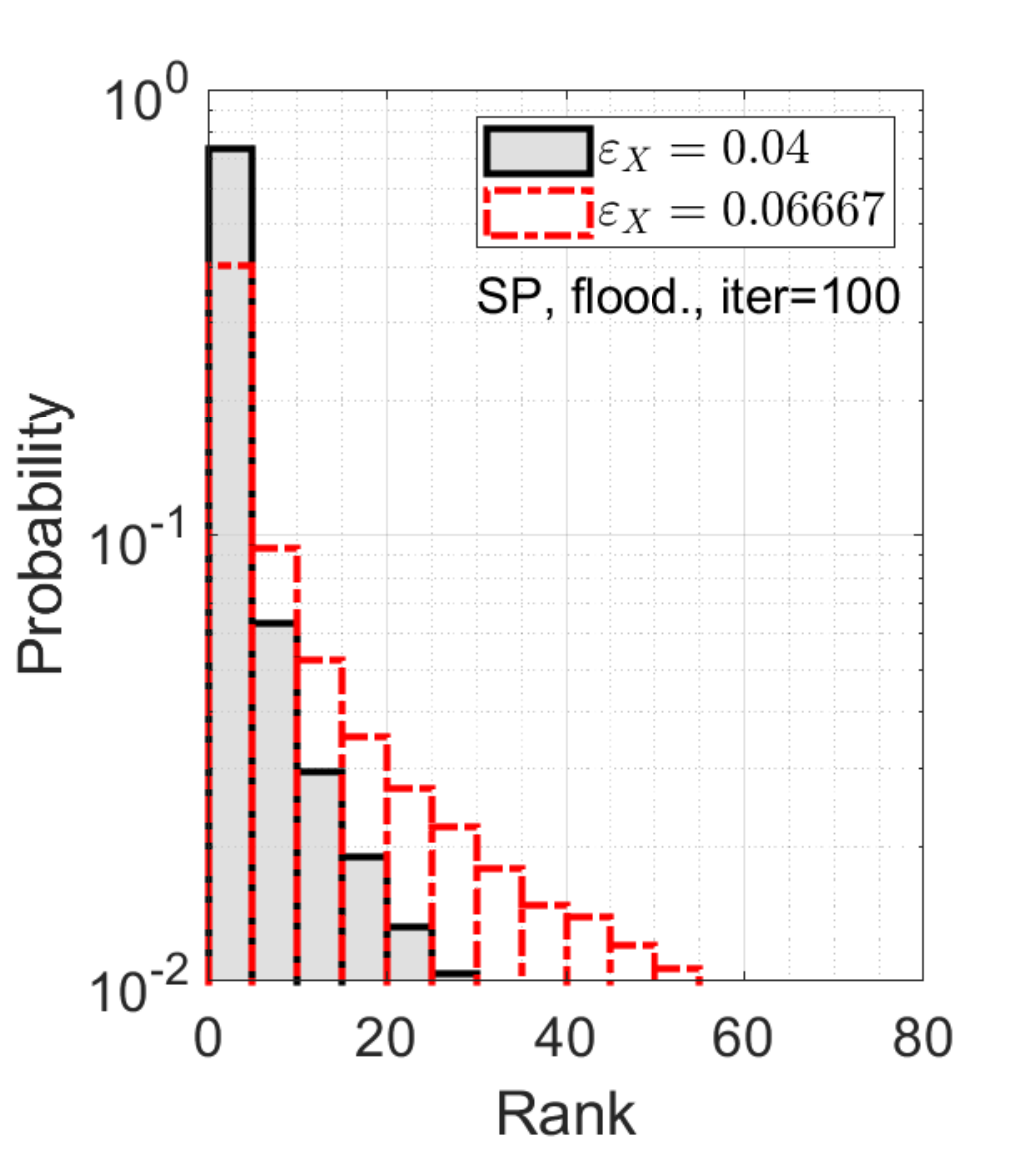}\qquad%
\includegraphics[width=.4\linewidth]{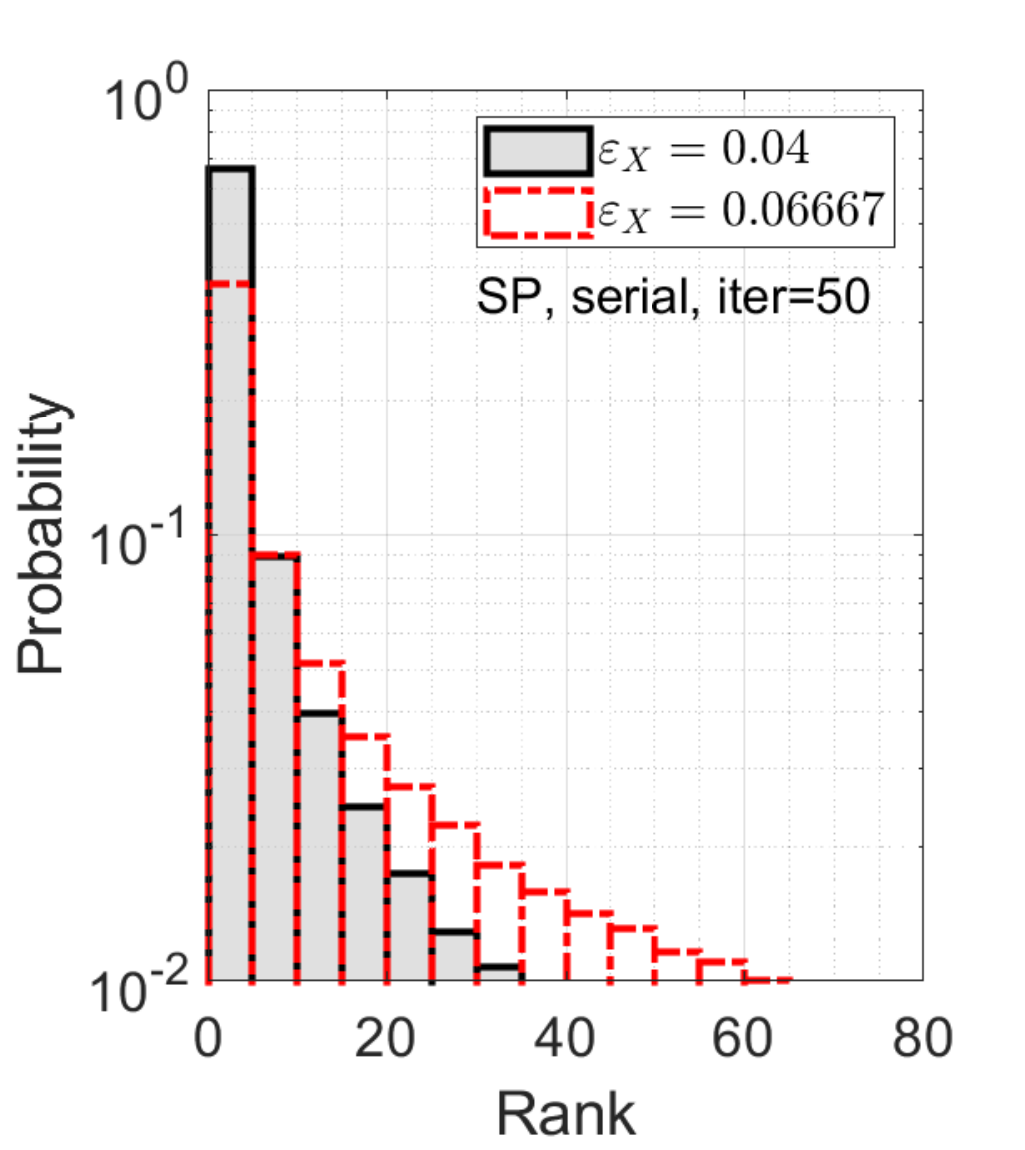}
\caption{Probability distribution of $\rank(r_X)$ values.}
\label{fig:Xchk_rank}
\end{figure}

Let $e_X$ be an $X$-error, and $s_X=H_Z e_X$ the corresponding syndrome. We discuss in this section the case where $e_X$ divides the support of some $X$-check $r_X$ in two equal parts, that is,  $X$ errors occur on exactly half of the qubits checked by $r_X$. 
For $e'_X := e_X + r_X$, we have  $H_Z e'_X = s_X$ and $|e'_X| = |e_X|$. Hence,  $e'_X$ and $e_X$ are  errors of the same weight,  differing on only $|r_Z|$ qubits, and which are valid (and equivalent) corrections of the syndrome. We  refer to such an error as a stabilizer-splitting error (see also the example in \cite[Fig.~2]{poulin2008iterative} and the \emph{symmetric stabilizer trapping sets} defined in \cite{raveendran2021trappingsetsof} for a more in-depth comprehension of the phenomenon). 
If an MP decoder is run with input syndrome $s_X$, it is drawn by $e'_X$ and $e_X$ in two different directions, with similar intensity. This may cause the MP decoder getting lost, while trying to find its way to a valid correction of the syndrome. 

Let $\left(\tilde{\gamma}_i\right)_{i}$ be the output soft information of the MP decoder, and $\hat{e}_X$ be the corresponding hard decision estimate (Eq.~(\ref{eq:hard-decision})). We define
\begin{equation}
\label{eq:delta_eX_eXp}
\displaystyle \delta(e_X, e'_X) \ := \sum_{i\in\supp(\hat{e}_X + e_X)} \!\!\!\! |\tilde{\gamma}_i| \ \ - \sum_{i\in\supp(\hat{e}_X + e'_X)} \!\!\! |\tilde{\gamma}_i|.
\end{equation}
Since flipping a hard-decision estimate value requires the corresponding soft information to change its sign, the two sums  in Eq.~(\ref{eq:delta_eX_eXp}) indicate the necessary change in the soft information, so that the corresponding hard decision estimate moves from $\hat{e}_X$ to either $e_X$ (left sum), or $e'_X$ (right sum). In case the MP decoder fails, that is $H_Z \hat{e}_X \neq s_X$, we expect that this is due to the decoder being attracted to a similar degree towards both $e_X$ and $e'_X$ (``lost in-between'' behavior). Put differently, we expect the $\delta(e_X, e'_X)$ value to be small.

Using $|\tilde{\gamma}_i| = (-1)^{\hat{e}_X(i)}\tilde{\gamma}_i$ and $e'_X = e_X + r_X$, one can easily verify that $\delta(e_X, e'_X)$ rewrites as
$ \delta(e_X, e'_X) = \sum_{i\in\supp(e_X)} \tilde{\gamma}_i - \sum_{i\in\supp(e'_X)} \tilde{\gamma}_i 
  = \sum_{i\in\supp(r_X)} (-1)^{e'_X(i)} \tilde{\gamma}_i$,
%
and therefore,
\begin{equation}
|\delta(e_X, e'_X)| \leq \tilde{\gamma}(r_X) := \sum_{i\in\supp(r_X)} | \tilde{\gamma}_i|
\end{equation}
We refer to $\tilde{\gamma}(r_X)$ as the reliability of the $X$-check $r_X$. Hence, we may use $X$-check reliability values to determine which $X$-checks are possibly responsible of a ``lost in-between'' behavior. To illustrate this, let us define
\begin{equation}
\rank(r_X) := |\{ r'_X \mid \tilde{\gamma}(r'_X) < \tilde{\gamma}(r_X)\}|.
\end{equation}
We generate random errors $e_X$, where each error divides the support of a random $X$-check $r_X$ in two equal parts, and compute the $\rank(r_X)$ values. Fig.~\ref{fig:Xchk_rank} shows the estimated probability distribution (histogram format) of $\rank(r_X)$ values. We consider the lifted product code B1$[882, 24]$ from~\cite{panteleev2021degenerate}, decoded by the SP algorithm, with both flooded and serial schedulings (we use 50 decoding iterations for serial scheduling, and 100 decoding iterations for flooded scheduling). We assume a depolarizing channel with $p_X = p_Y = p_Z = p/3$, where $p=0.1$ or $p=0.06$, giving $\varepsilon_X = 0.0667$ or $\varepsilon_X = 0.04$ (Section~\ref{subsec:pauli_errors}). As expected, it can be observed that the $\rank(r_X)$ value is small with high probability (the width of histogram bins is equal to $5$).

\subsection{SI Post-Processing}

We consider the situation where the MP decoder fails to find a hard decision estimate $\hat{e}_X$ satisfying the given syndrome $s_X$. The idea of the SI post-processing (Algorithm~\ref{alg:SI-post-processing}) is to {\em inactivate} the qubits in the support of the less reliable $X$-checks. By a slight abuse of language we shall say that we inactivate the $X$-check itself (rather than its support).  We start by inactivating the least reliable $X$-check, and let this be $r_X$. This means that we discard the qubits checked by $r_X$ from the MP decoding process. Precisely, let
\begin{equation}
\label{eq:HZ_system}
H_Z = \left[ \begin{array}{cc}
H_{Z | r_X} & A \\
0 & H_{Z | \overline{r_X}}
\end{array} \right],
\end{equation}
where $H_{Z | r_X}$ is the submatrix determined by the columns corresponding to the qubits in $\supp(r_X)$, and the rows having at least one non-zero entry in any one of these columns\footnote{Note that $r_X \in \ker H_Z$, hence any row of $H_Z$ has an even number ($\geq 0$) of non-zero entries in the columns corresponding to $\supp(r_X)$.} (for the sake of illustration, in~(\ref{eq:HZ_system}) we assumed that $H_{Z | r_X}$ is a leading submatrix). Then we rerun  MP decoding on the matrix $H_{Z | \overline{r_X}}$, with input syndrome $s_{X | \overline{r_X}} = [0\ H_{Z | \overline{r_X}}] e_X$ (that is, syndrome $s_X$ is restricted to the rows of the $H_{Z | \overline{r_X}}$ submatrix only). If MP decoding succeeds, it provides an estimate $\hat{e}_{X | \overline{r_X}}$ of the error  outside $\supp(r_X)$. To estimate the error  on $\supp(r_X)$, we solve the  linear system $H_{Z | r_X} e_{X | r_X} = s_{X | r_X} + A \hat{e}_{X | \overline{r_X}}$. If the system has a solution\footnote{Note that the system will have at least two solutions. We pick any of them.}, say $\hat{e}_{X | {r}_X}$, the decoding process stops and outputs $\hat{e}_X := (\hat{e}_{X | {r}_X}, \hat{e}_{X | \overline{r_X}})$. In case that either the system has no solution, or the MP decoding fails, SI post-processing continues by inactivating the next least reliable $X$-check, until a maximum (fixed) number of $X$-check inactivations, denoted by $\lambda_\text{max}$, is reached.

Two observations are in place here: (i) $X$-check reliability values are {\em computed only once, after the initial MP decoding attempt} (run on the whole $H_Z$ matrix). (ii) $X$-checks are sorted in increasing order of  reliability, and  inactivated, {\em one at a time}, in this order.  In our experiments, run on qLDPC codes of length less than $2000$ qubits, inactivating several $X$-checks at the same time did not improve the decoding performance. 


\begin{algorithm}[!t]
\caption{X-error SI$[\lambda_\text{max}]$ post-processing}
\label{alg:SI-post-processing}
\DontPrintSemicolon

  $\hat{e}_X \leftarrow \text{MP}(H_Z,s_X)$ \;
  \If{$ H_Z \hat{e}_X = s_X$}
   {  		
   		Return $\hat{e}_X$\;
	} 
	\Else{
	Compute $X$-check reliability values \; 
	\hspace*{7mm} $\tilde{\gamma}(r_X) = \sum_{i\in \supp(r_X)} \left| \tilde{\gamma}_i \right|,\  \forall  r_X$ \;
	Sort $X$-checks   in increasing order of reliability \;
	
	\smallskip
	\For{$1\leq \lambda \leq \lambda_\text{\rm max}$}
   {
   		$r_X  \leftarrow$ next least reliable $X$-check \;
   		
	    $\hat{e}_{X | \overline{r_X}} \leftarrow \text{MP}\left(H_{Z | \overline{r_X}}, s_{X | \overline{r_X}}\right)$ \; 
	    
	    \smallskip  		
   		\If{$H_{Z | \overline{r_X}X} \hat{e}_{X | \overline{r_X}} \neq s_{X | \overline{r_X}}$}
   		{
   			Continue\;
   		}
   		\Else{
   		Solve $H_{Z | r_X} e_{X | r_X} = s_{X | r_X} + A \hat{e}_{X | \overline{r_X}}$\;
   		\If{the system has a solution $\hat{e}_{X | {r}_X}$}
   		{
   			Return $\hat{e}_X := (\hat{e}_{X | {r}_X}, \hat{e}_{X | \overline{r_X}})$\;
   		}
   		\Else
   		{
   			Continue\;
   		}
   		}
   		
   }
   }
   \smallskip
   Return decoding failure \:
\end{algorithm}  

Finally, as mentioned above, the following two situations may occur, causing SI post-processing to continue: (i) the MP decoder does not converge to a solution on the reduced matrix $H_{Z | \overline{r_X}}$, or (ii) it does converge, but the system on the remaining qubits cannot be solved. The first kind of error typically happens when the inactivated $X$-check  was not one that implied the failure in the first place (initial MP decoding attempt on the full $H_Z$ matrix). The second kind of error happens when the MP decoding converged to a wrong solution on the reduced matrix $H_{Z | \overline{r_X}}$. This may happen when some of the active qubits become ``dead ends'' (degree $1$ is the reduced matrix $H_{Z | \overline{r_X}}$), thus ``absorbing'' the error around the $Z$-checks they are connected to.


\subsection{Complexity}

The SI$[\lambda_\text{max}]$ post-processing step has worst-case complexity $\mathcal{O}(\lambda_\text{max} n \log n)$, where $\lambda_\text{max}$ is the maximum number of inactivated $X$-checks and $\mathcal{O}(n \log n)$ is the complexity of the MP decoding (assuming the number of decoding iterations increases as $\log n$~\cite{savin2014ldpc}). Solving the  linear system in the SI post-processing step has constant complexity, since the size of the system does not exceed the maximum $X$-check weight. In case $\lambda_\text{max}$ is not a constant, but scales linearly with $n$, the worst-case complexity  becomes  $\mathcal{O}(n^2 \log n)$. In any case, the average-case complexity is $\mathcal{O}(\lambda_\text{ave} n \log n)$, where $\lambda_\text{ave}$ is the average number of inactivated $X$-checks during SI post-processing.

The OSD-0 post-processing needs to solve a linear system, whose size scales linearly with $n$. Thus, its complexity is at most $\mathcal{O}(n^3)$. The best algorithm to solve a general system of linear equations has complexity $\mathcal{O}(n^{2.38})$~\cite{golub2013matrix}, although it is of little practical use (except for extremely large systems). 


\section{Numerical Results}
\subsection{Logical Error Rate Performance}
\label{sec:Logical Error Rate Performance}
\begin{figure*}[h!]
\centering
\includegraphics[width=.48\linewidth]{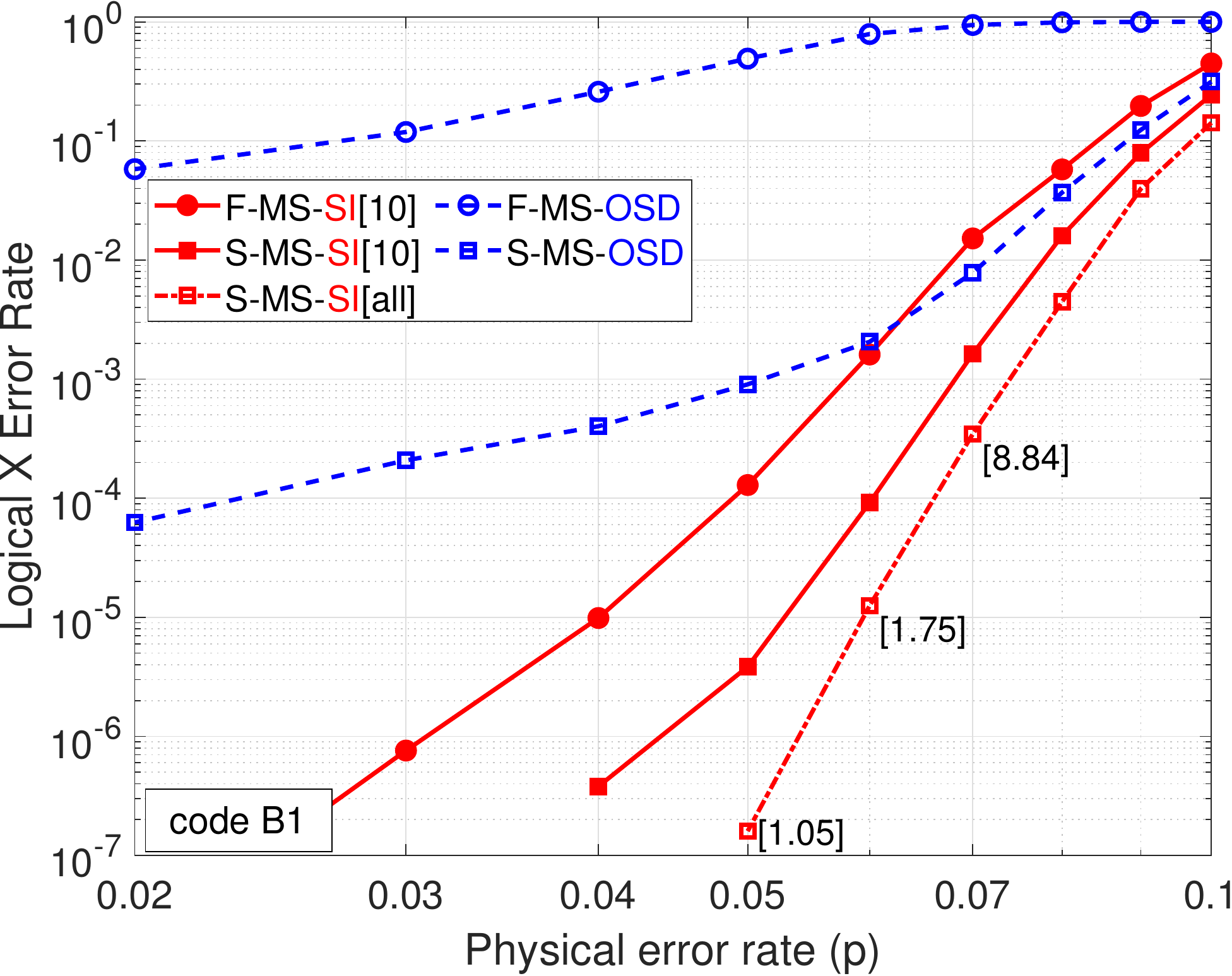}\hfill%
\includegraphics[width=.48\linewidth]{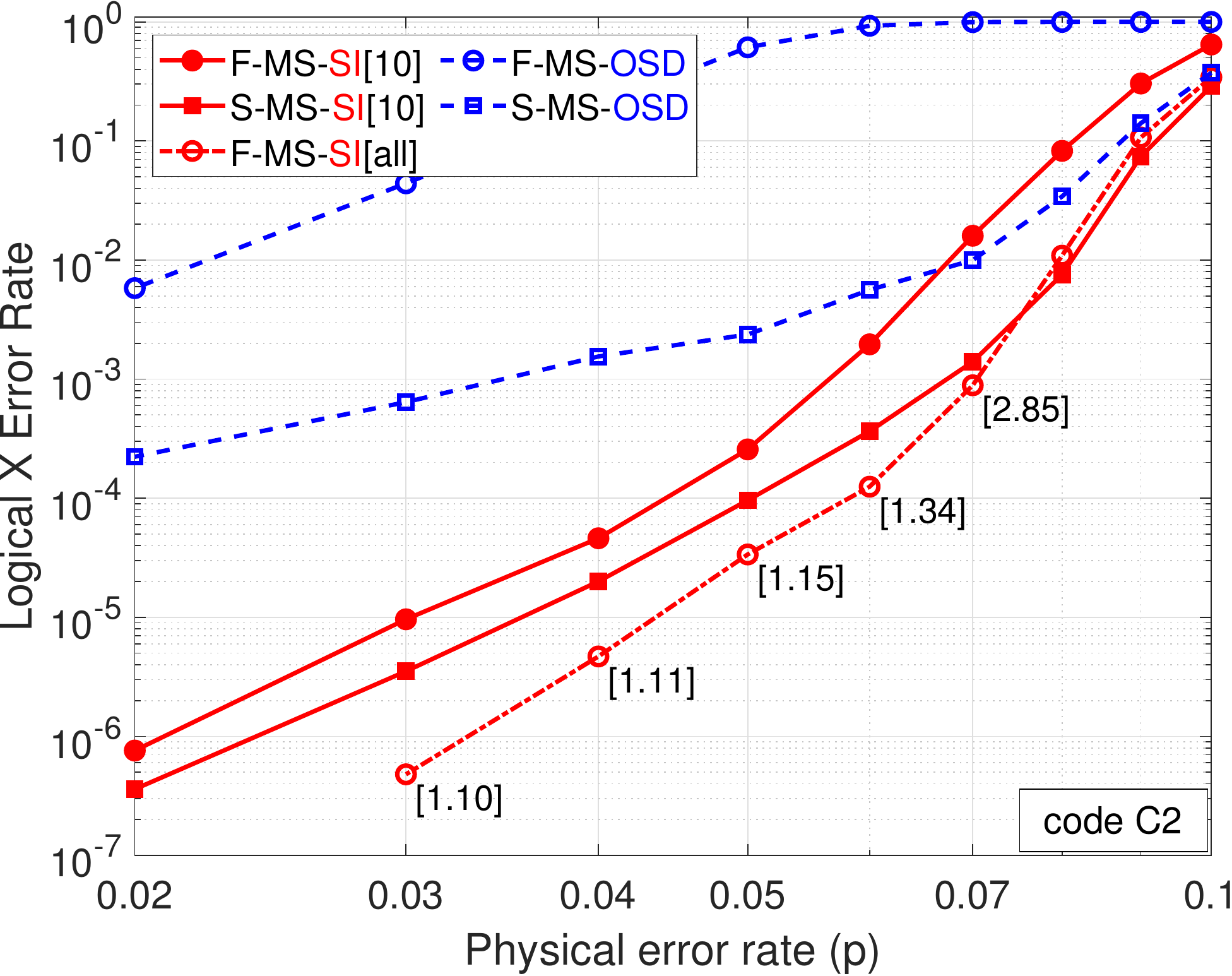}\\[2mm]
\includegraphics[width=.48\linewidth]{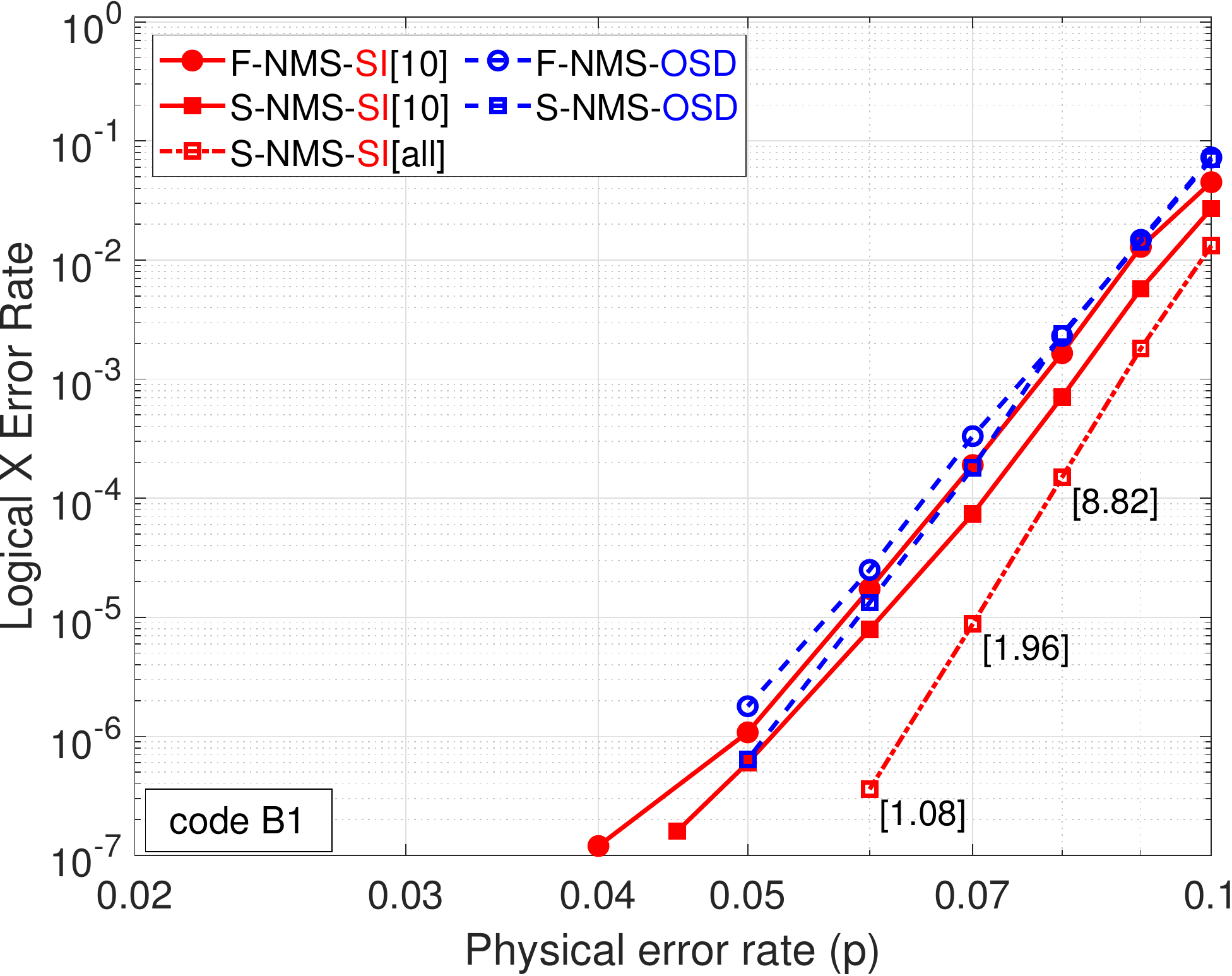}\hfill%
\includegraphics[width=.48\linewidth]{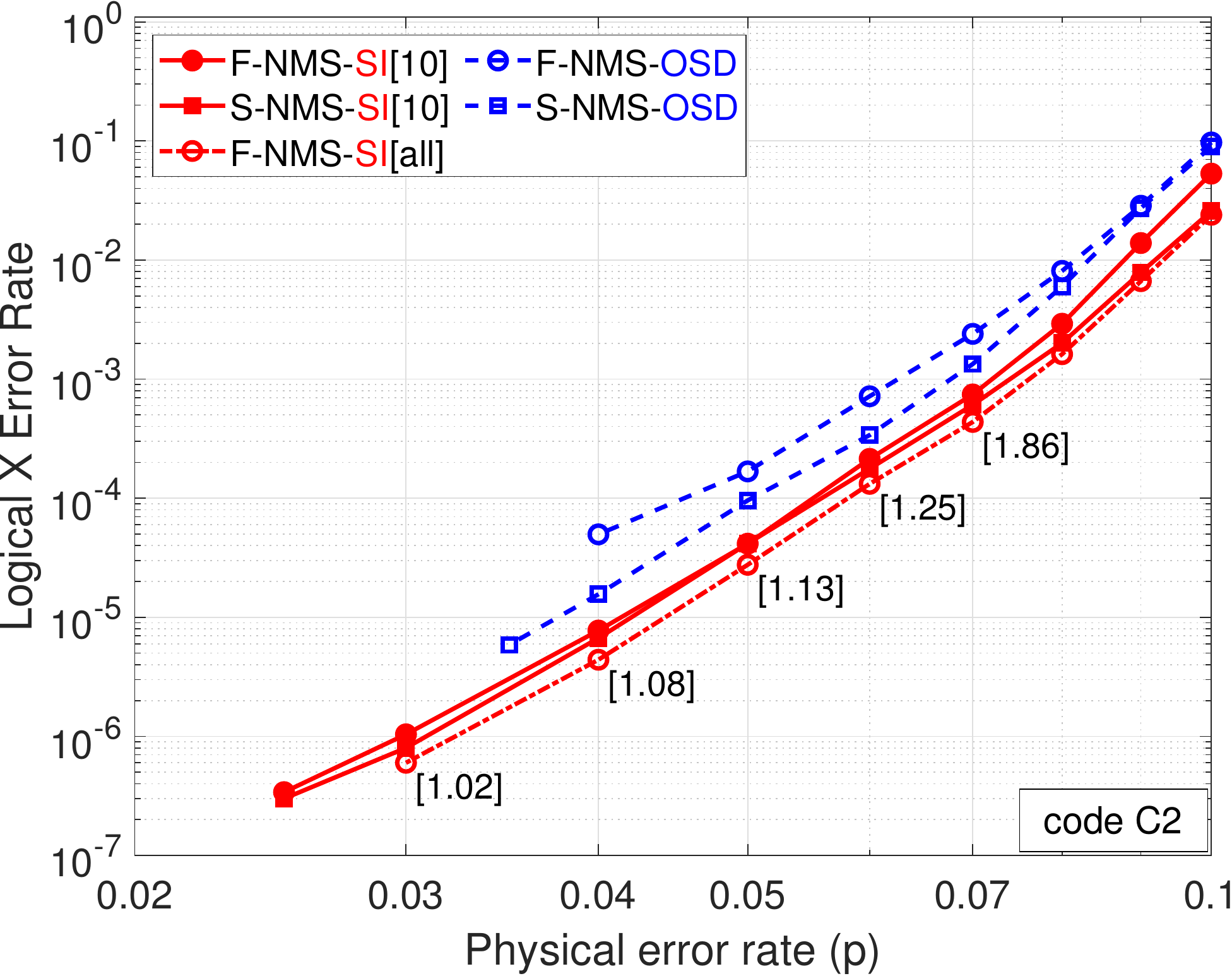}\\[2mm]
\includegraphics[width=.48\linewidth]{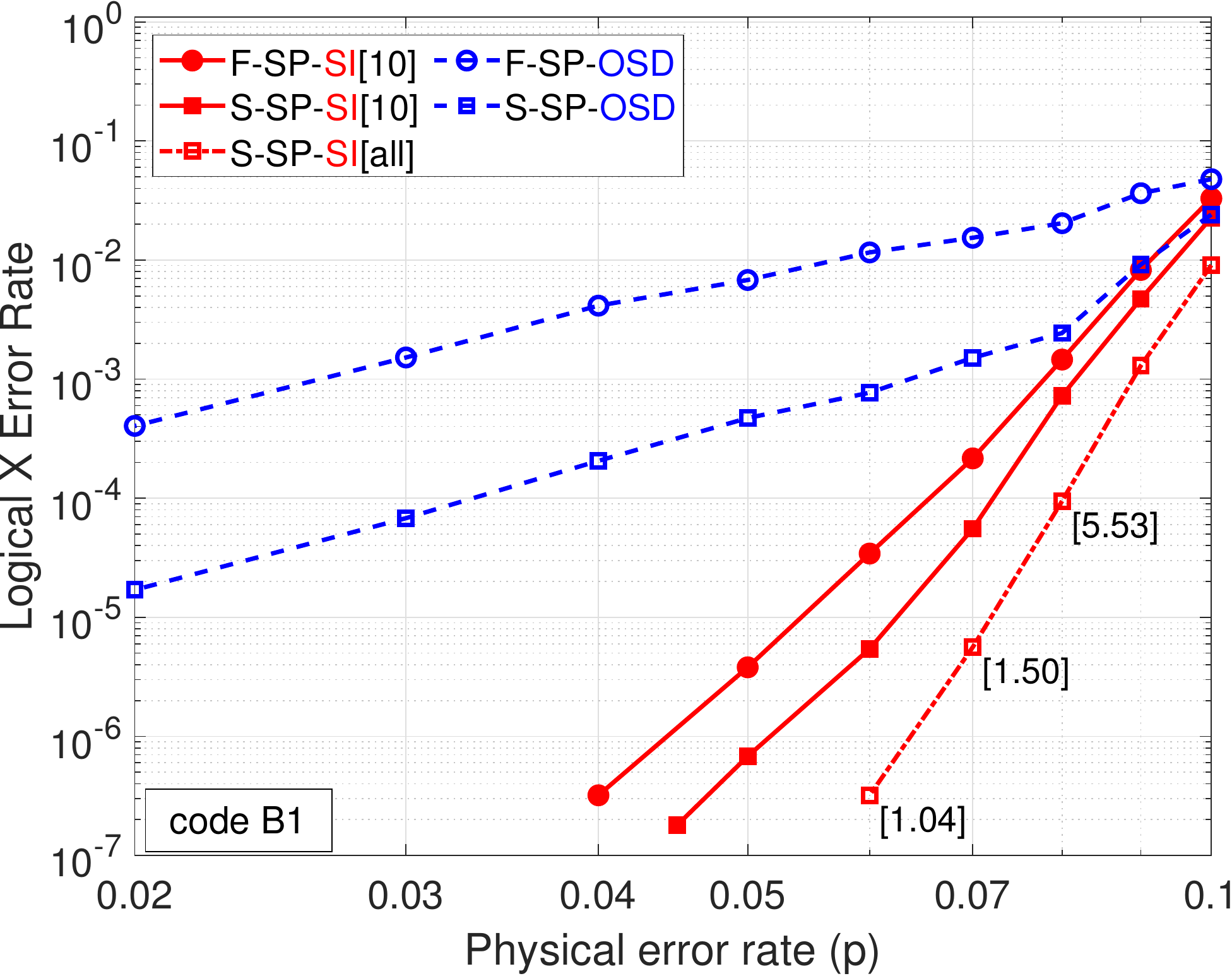}\hfill%
\includegraphics[width=.48\linewidth]{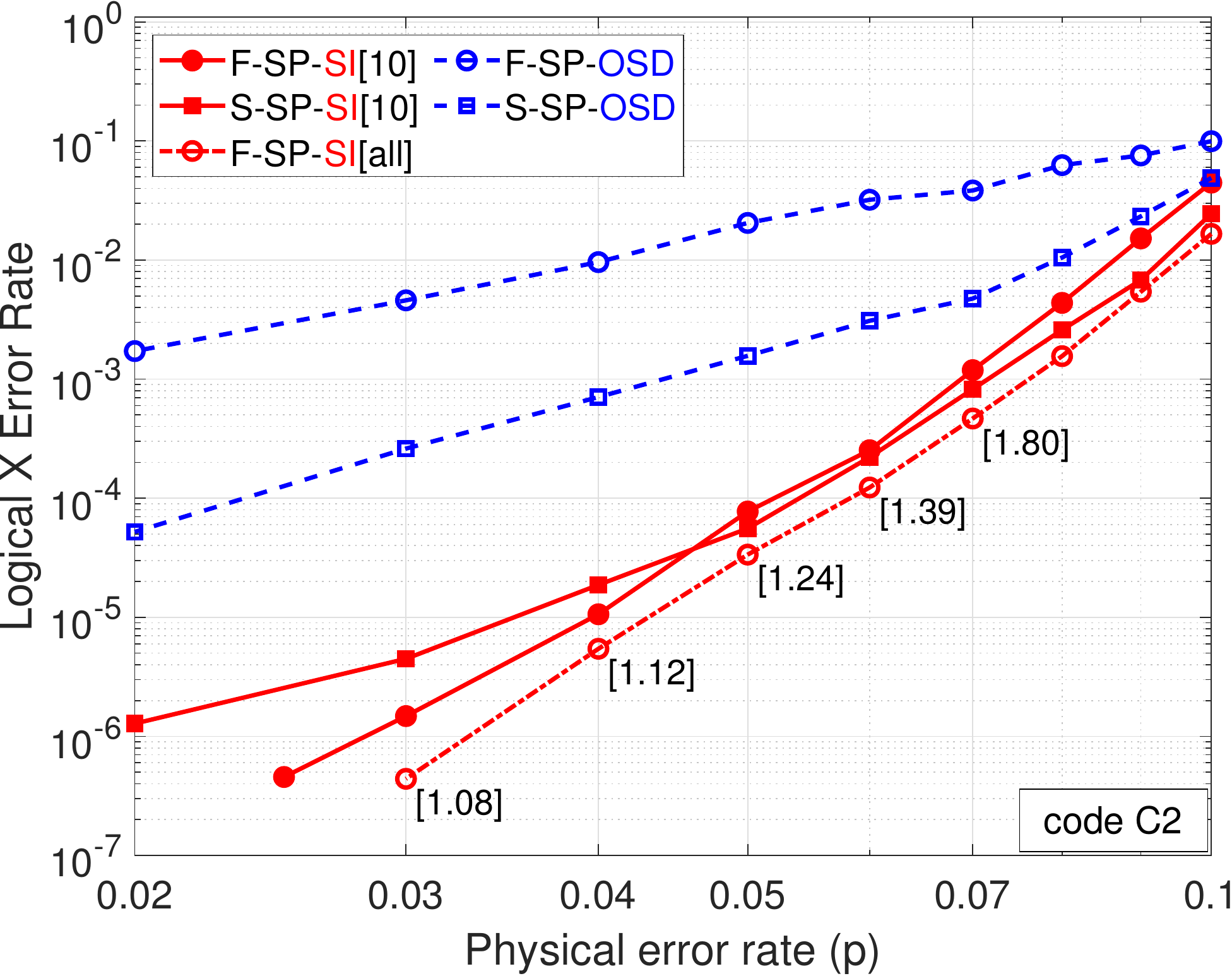}
\caption{Comparison of SI and OSD on B1$[882,24,\leq 24]$ code (left) and C2$[1922,50,16]$ code (right) with MS (top), NMS (middle) and SP (bottom) decoders and flooded (`F', circle markers) and serial (`S', square markers) scheduling. OSD post-processing is of order $0$. For NMS, we use normalization factor 0.625 for OSD (following \cite{panteleev2021degenerate}) and 0.9 for SI. All MP decoders use 50 decoding iterations for serial scheduling, and 100 decoding iterations for flooded scheduling.}
\label{fig:num_res}
\end{figure*}

In Fig.~\ref{fig:num_res}, we provide numerical results for the codes B1$[881,24,\leq 24]$ and C2$[1922,50,16]$ from~\cite{panteleev2021degenerate}. The first is a lifted product code, and the second a hypergraph-product code. We note that for both B1 and C2 codes, the constituent classical LDPC codes have column weight at least $3$, and no cycles of length $4$. These characteristics are  well suited to the proposed SI post-processing method, since except for the resolution of a small, constant size, linear system, it entirely relies on MP decoding. 

\smallskip We consider the depolarizing channel, with physical error rate $p$ (shown on the abscissa), and $p_X = p_y = p_Z = p/3$. Fig.~\ref{fig:num_res} shows the logical-X error rate, for three different MP algorithms, and two different schedulings, using either SI or OSD-0 post-processing.  It can be observed that SI$[\lambda_\text{max}=10]$ significantly outperforms the OSD-0 post-processing\footnote{We note that we perform OSD-0 post-processing on binary MP decoding (of $X$-errors), which explains the slight shift of the curves with respect to \cite{panteleev2021degenerate}, where  non-binary MP decoding (of both $X$ and $Z$ errors) is performed.} in case of either MS or SP decoding, for both scheduling strategies. For NMS decoding\footnote{For the NMS decoding, the normalization factor is chosen so as to ensure the best performance of the corresponding post-processing step.  NMS-OSD exhibits very good performance using normalization factor $0.625$ (the same normalisation factor was used in~\cite{panteleev2021degenerate}, for non-binary NMS decoding). However, its performance would have been severely degraded, if the normalization factor had been set to $0.9$. Conversely, NMS-SI performs well when the normalization factor is set to $0.9$, but rather poor when it is set to $0.625$.},  the performance of SI$[\lambda_\text{max}=10]$ is only slightly better than that of OSD-0, but comes however at the cost of a significantly reduced complexity. 

\smallskip  The SI[all] curves in Fig.~\ref{fig:num_res} correspond to the case when the $\lambda_{\text{max}}$ value is set to the total number of $X$-checks  (number of rows of $H_X$), thus providing the best achievable performance by the proposed SI post-processing.  (We note that for the code B1, the SI[all] performance is slightly better using serial scheduling than using flooding scheduling, while the opposite occurs for the code C2 -- not shown in the figure). Finally, in order to evaluate the average-case complexity, we also show in Fig.~\ref{fig:num_res} the $\lambda_\text{ave}$ values for the SI[all] post-processing, in the lower part of the waterfall region (logical error rates less than $10^{-3}$). In order to avoid a bias due to MP decoding performance, the average is computed only over the cases when the SI post-processing is used (MP decoding failed), thus $1 \leq \lambda_\text{ave} \leq \lambda_\text{max}$. It can be seen that $\leq \lambda_\text{ave}$ approaches $1$ for  low error probabilities, giving an average-case complexity slightly higher than the MP decoding complexity. 




\clearpage

\subsection{Threshold}
\begin{figure*}[!t]
\centering
\subfigure[{NMS-SI$[\lambda_\text{max}=0.02\,m_X]$}]{\includegraphics[width=.49\linewidth]{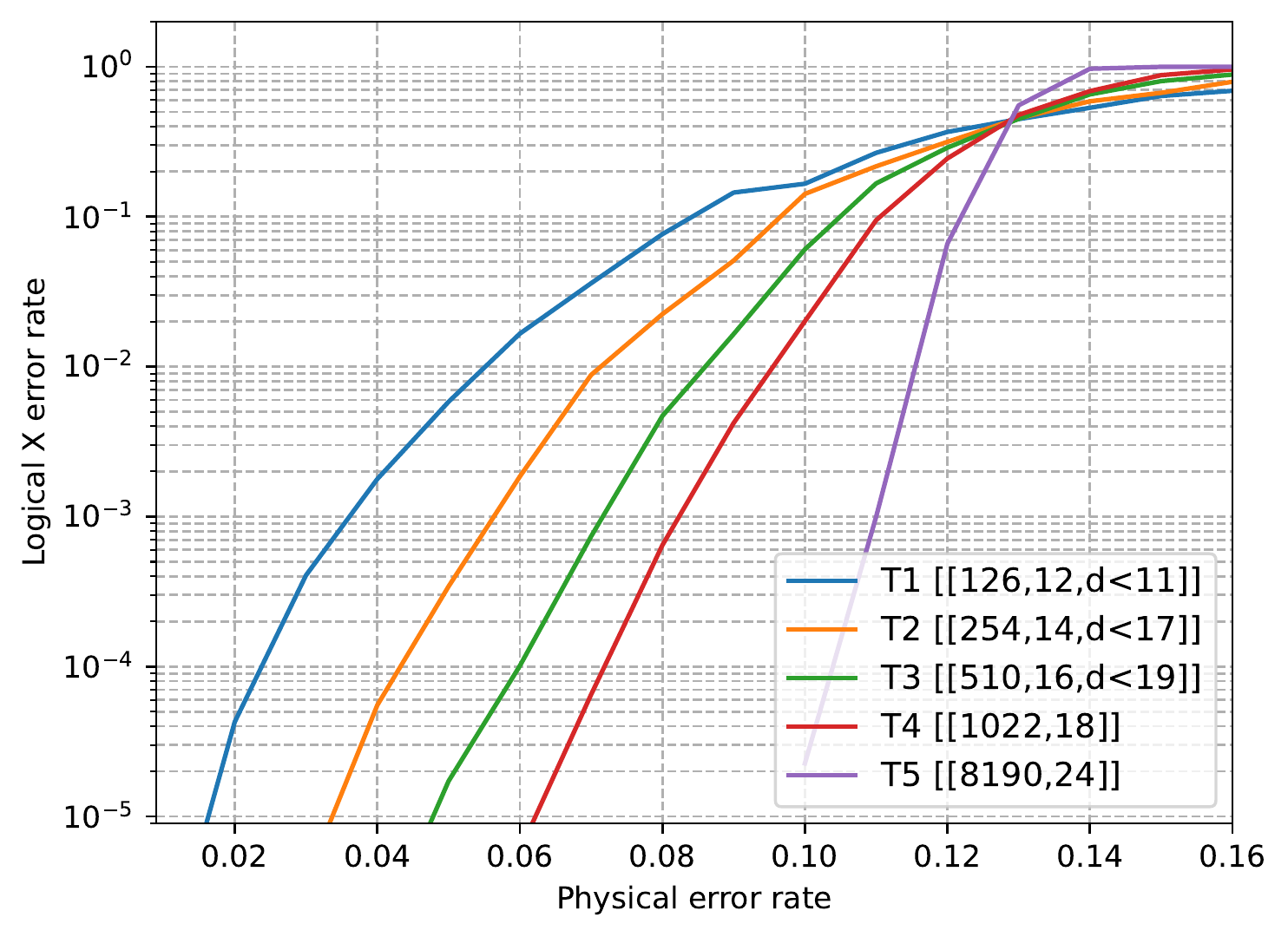}}\hfill%
\subfigure[NMS-OSD-0]{\includegraphics[width=.49\linewidth]{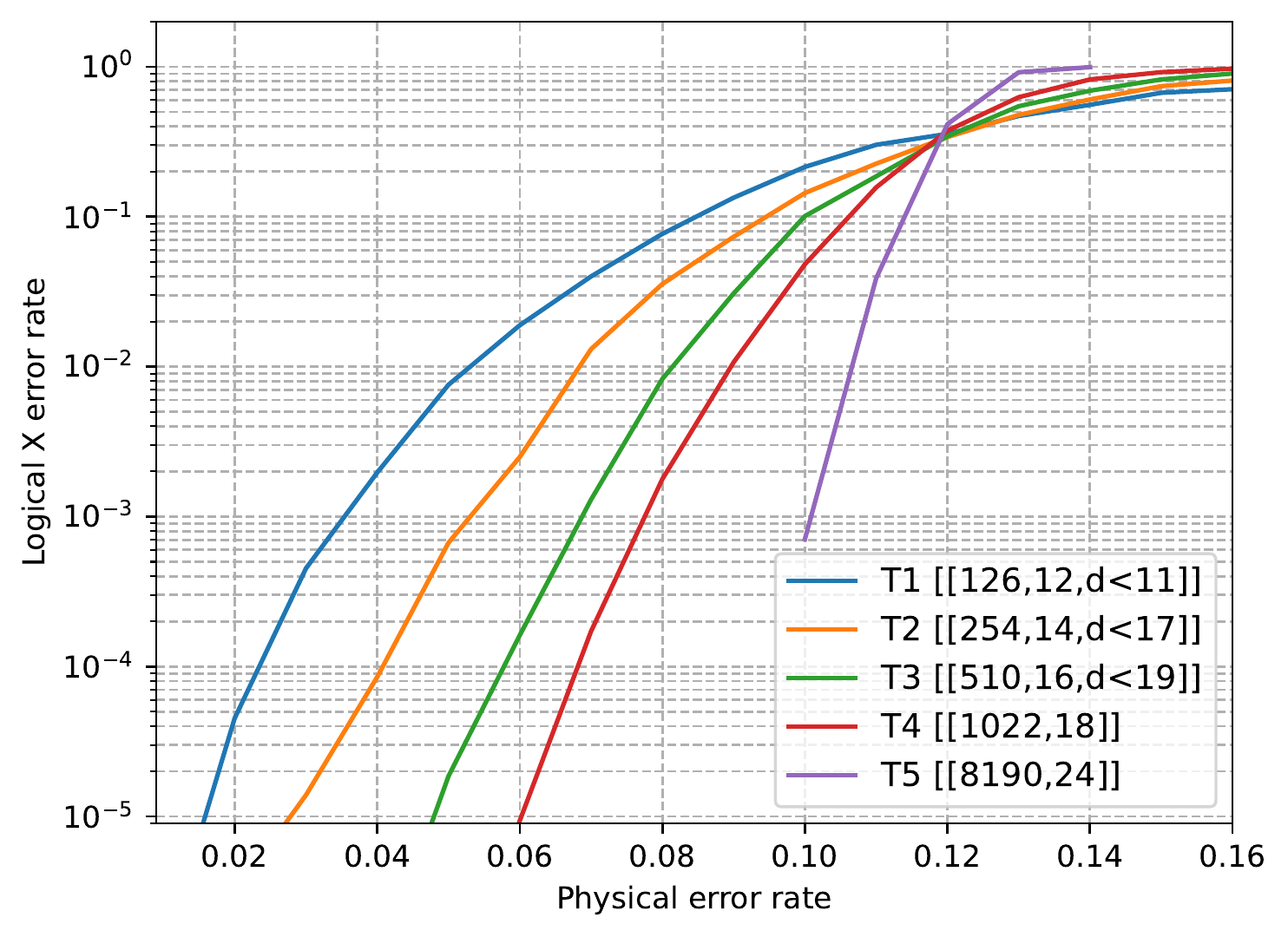}}
\caption{Threshold for SI$[\lambda_\text{max}=0.02\,m_X]$ and OSD-0 on a family of generalized bicycle codes with NMS and serial scheduling. For NMS, we use normalization factor 0.625 for OSD (following \cite{panteleev2021degenerate}) and 0.9 for SI. All MP decoders use serial scheduling with at most 100 decoding iterations.  
}
\label{fig:thresh}
\end{figure*}

{
In Fig.~\ref{fig:thresh}, we provide numerical evidence that SI can achieve a threshold for a family of generalized bicycle codes\footnote{ These codes can be found in the alist format at \url{https://gricad-gitlab.univ-grenoble-alpes.fr/ducrestj/qldpc-codes}} proposed in \cite[Appendix C, Fig.\,9]{panteleev2021degenerate}. 
 Generalized bicycle codes (first studied in \cite{kovalev2013quantum}) are constructed by using two commuting square matrices $A$ and $B$, with $H_X = [A,B],\quad H_Z = [B^T,A^T]$. For the family of codes considered here, $A$ and $B$ are circulant matrices, thus they  commute, and code parameters are given by $[[n,k]] = [[2^{s+1}-2, 2s]]$, for $s=6,7,8,9,12$.
 
 
\smallskip We estimate the threshold under MP decoding, with both SI and OSD post-processing, for  the depolarizing channel (same channel model as before). For a fair comparison between SI and OSD, we use NMS with appropriate normalization factors. The scheduling chosen is serial, as it allows faster convergence and better results in most cases. 

\smallskip  For the SI post-processing, we choose $\lambda_\text{max} = 10$ for the code of length $1022$ qubits, representing 2\% of the number of $X$-checks, denoted by $m_X$. Note that $m_X = (n-k)/2$. Then, we keep the same ratio between the maximum number of inactivated $X$-checks, and the total number of $X$-checks, that is, we consider $\lambda_\text{max} = 0.02\,m_X$, for  the five simulated codes\footnote{ Precisely, $\lambda_\text{max} = 2, 3, 5, 10, 82$, for code length $n=126, 254, 510, 1022, 8190$, respectively.}.  Since $m_X \approx n/2$, the worst case complexity of the MP-SI decoder scales as $0.01 n^2 \log(n)$.  However, we note that the average case complexity approaches $O(n\log n)$ in the low error rate region (that is, $\lambda_\text{ave}$ approaches $1$ for  low error probabilities, similar to the observation made in Section~\ref{sec:Logical Error Rate Performance}).

\smallskip The threshold given by SI is around 13\%, while the one of OSD-0 is only 12\%. Apart from the numerical results, the computational advantage of SI becomes clear for codes above a thousand qubits. The observed threshold phenomenon shows that the SI post-processing scales well with longer codes, in spite of the fact that it inactivates only one check  at a time, and the inactivation is tested for quite a small (2\%) fraction of checks.
 
 } 

\section{Conclusion and Perspectives}
The degeneracy of qLPDC codes is 
the main reason for the inefficiency of MP decoders. To cope with this phenomenon, the SI post-processing method inactivates a set of qubits, supporting a check in the dual code.  Except for the resolution of a small, constant size, linear system, the proposed MP-SI approach entirely relies on MP decoding. Its low complexity and compatibility with various MP decoding algorithms and scheduling strategies make it an attractive and practical decoding solution. 
Our numerical simulations showed that MP-SI provides effective results, especially when the constituent classical LDPC codes have column weight at least $3$, and no cycles of length $4$.  {
We have also shown that SI post-processing can achieve a threshold on a family of generalized bicycle codes, outperforming the one achieved by OSD. 

\smallskip Finally, we note that in this work we considered binary MP decoding algorithms, where $X$ and $Z$ errors are decoded separately. While correlation between $X$ and $Z$ errors can still be taken into account by binary decoding (\emph{e.g.}, by initializing the $Z$-error decoding conditional on the output of the $X$-error decoding), a more effective approach to deal with correlated errors consists of the use non-binary decoding. Since the state of the art for correlated Pauli errors remains the non-binary NMS decoder with OSD post-processing~\cite{panteleev2021degenerate}, it would be interesting to generalize the proposed SI post-processing to the non-binary case.
}



\section*{Acknowledgment}
This work was supported by the QuantERA grant EQUIP, by the French Agence Nationale de la Recherche, ANR-22-QUA2-0005-01, and by the Plan France 2030 through the project ANR-22-PETQ-0006.

\bibliographystyle{IEEEtran}
\bibliography{biblio_database}

\end{document}